\pdfoutput=1

\documentclass[namedreferences]{solarphysics}

\usepackage[hyperref,optionalrh]{spr-sola-addons} % For Solar Physics
\usepackage{graphicx}        % For eps figures, newer & more powerfull
\usepackage{color}           % For color text: \color command
\usepackage{breakurl}        % For breaking URLs easily trough lines
\newcommand{\etal}{{\it et al.}}

\newcommand{\aap}{    {\it Astron. Astrophys.}}

\newcommand{\apj}{    {\it Astrophys. J.}}
\newcommand{\apjl}{   {\it Astrophys. J. Lett.}}

\newcommand{\jgr}{    {\it J. Geophys. Res.}}

\newcommand{\solphys}{{\it Solar Phys.}}
\newcommand{\sovast}{ {\it Soviet  Astron.}}

\chardef\us=`\_

\begin{document}

\begin{article}

\begin{opening}

\title{New Types of the Chromospheric Anemone Microflares:
       Case Study}

\author[addressref={GAISh,IKI},corref,email={dumin@sai.msu.ru}]%
  {\fnm{Yurii~V.}~\lnm{Dumin}\orcid{0000-0002-9604-5304}}%\sep
\author[addressref=GAISh,email={somov-boris@mail.ru}]%
  {\fnm{Boris~V.}~\lnm{Somov}}%\sep
%\author{\inits{}\fnm{}~\lnm{}\orcid{}}

\address[id=GAISh]{
P.K.~Sternberg Astronomical Institute (GAISh) of
M.V.~Lomonosov Moscow State University,
Universitetskii prosp.\ 13, 119234, Moscow, Russia}
\address[id=IKI]{
Space Research Institute (IKI) of Russian Academy of Sciences,
Profsoyuznaya str.\ 84/32, 117997, Moscow, Russia}

\runningauthor{Yu.V.~Dumin and B.V.~Somov}
\runningtitle{New Types of the Chromospheric Anemone Microflares}

\begin{abstract}
The chromospheric anemone microflares (AMF) are the transient solar phenomena
whose emission regions have a multi-ribbon configuration.
As distinct from the so-called ``atypical'' solar flares, also possessing
a few ribbons, the temporal and spatial scales of AMFs are a few times less,
and the configuration of their ribbons is more specific (star-like).
The previously reported AMFs had typically three or, less frequently, four
ribbons; and it was shown in our recent paper (Dumin and Somov: 2019,{\aap}
\textbf{623}, L4) that they can be reasonably described by the so-called
GKSS model of magnetic field, involving as few as four point-like magnetic
sources with various polarity and arrangement.
To seek for the new types of AMFs, we performed inspection of the large
set of the emission patterns in the chromospheric line~{Ca}\,{\small II}\,H
recorded by \textit{Hinode}/SOT and confronted them with the respective
magnetograms obtained by \textit{SDO}/HMI.
As follows from this analysis, it is really possible to identify the new
unusual AMFs.
Firstly, these are the flares occurring in the regions with unbalanced
magnetic flux.
Secondly, and most interesting, it is possible to identify the AMFs with
much more complex spatial configurations, \textit{e.g.}, involving five
luminous ribbons with a nontrivial arrangement.
As follows from the corresponding magnetograms, they are produced by
the effective magnetic sources (sunspots) of different polarity with
intermittent arrangement, but their number is greater than in the standard
GKSS model.
\end{abstract}
\keywords{Flares, Microflares and Nanoflares;
          Flares, Relation to Magnetic Field}

\end{opening}

\section{Introduction}
\label{sec:Intro}

The so-called anemone microflares (AMF) are the specific type of the
small-scale flaring phenomena in the solar atmosphere, which were discovered
in the chromospheric line~{Ca}\,{\small II}\,H by the Solar Optical Telescope
(SOT) onboard \textit{Hinode} satellite soon after beginning of its operation
\citep{Shibata_07}.
These flares are characterized by the unusual spatial configuration of
the emission regions formed at the initial stage of their development:
as distinct from two ribbons in the ordinary large solar flares, the AMFs
possess three or, less frequently, four ribbons oriented at various angles
with respect to each other.%
\footnote{%
Such microflares should not be mixed with the specific large active regions
and flares that sometimes are also called the anemone \citep{Asai_09}.}
At the second stage, the AMFs are usually followed by the eruption upward,
forming the chromospheric jets \citep{Nishizuka_11,Singh_11,Singh_12}; but
this stage will not be discussed in more detail in the present paper,
because we study only the events occurring on disk.
A promising tool for observation of AMFs is the New Solar
Telescope (NST) in the Big Bear Solar Observatory, possessing the resolution
of~$ 0''\!.16 $ in He\,{\small I} 10\,830\,{\AA} line \citep{Zeng_16}, which
is substantially better than the resolution of \textit{Hinode}/SOT.

To avoid misunderstanding, let us emphasize that when we speak about
AMFs, we imply just the nontrivial emission patterns in the certain spectral
line.
The AMFs are definitely not an ``isolated'' physical entity: they might
manifest themselves also in other spectral ranges, \textit{e.g.}, EUV and
X-rays but with different geometric configurations.
However, the AMFs cannot be immediately identified with the majority of EUV
\citep{Harrison_97,Curdt_11} or X-ray microflares reported earlier
\citep{Christe_08,Hannah_08}, because the last-cited events possess the
spatial and temporal scales a few times greater than AMFs.
Nevertheless, the recent study by \citet{Zeng_16} revealed the EUV and X-ray
counterparts of the particular chromospheric microflare.

There are also the large solar flares possessing a few (more than two)
emissive ribbons, which were called the ``atypical'' multi-ribbon flares
\citep{Wang_14}; and some of them possess an extremely complex structure of
the ribbons, which are sophisticatedly entangled
\citep[\textit{e.g.}, middle panel of Figure~3 in][]{Dalmasse_15}.
However, one should not mix two above-mentioned types of the multi-ribbon
flares, because both their morphological characteristics and the magnetic field
structure are very different, as summarized in Table~\ref{table:Comparison}.
Besides, both the spatial and temporal scales of the atypical multi-ribbon
flares are a few times greater than for the anemone microflares.
There is also a clear qualitative difference in the orientation of the
emissive ribbons: quasi-parallel in the first case \textit{vs.} diverging
(star-like) in the second case.

At last, as regards the mechanism of their formation, emergence of a few
ribbons in the atypical flares is immediately associated with a few sets
of the ``fish-bone'' magnetic arcades, connecting the sunspots of opposite
polarity (typically, two sets of the arcades produce three emissive ribbons,
\textit{e.g.}, Figures~1(c,d) and 2(a,c) in \opencite{Wang_14}).
In other words, a few absolutely different magnetic fluxes are involved in
this process.
On the other hand, the diverging emissive ribbons in the anemone microflares
are caused most probably by the same magnetic flux split into a few magnetic
tubes at some height above the photosphere.
In other words, a few magnetic sources at the surface of photosphere produce
the magnetic fluxes experiencing a complex topological transformation and
interconnecting with each other at some height.

%%%%%%%%%%%%%%%%%%%%%%%%%%%%%%%%%%%%%%%%%%%%%%%%%%
\begin{table}
\caption{Comparison of the main characteristics of the ``atypical'' and
``anemone'' multi-ribbon flares.}
\label{table:Comparison}
\begin{tabular}{lcccc}
\hline
 && Atypical multi-ribbon && Anemone (star-like) \\
 && flares && microflares \\
\hline
Spatial size && $ 10''{-}\,25'' $ && $ {\approx}\,5'' $ \\
Lifetime && $ 15{-}20 $~min && $ 2{-}4 $~min \\[0.3ex]
Orientation of && approximately parallel && diverging in different \\[-0.2ex]
\quad the ribbons && to each other && directions \\[0.3ex]
Mechanism of && ``fish-bone'' sets of the && splitting of the magnetic
  \\[-0.2ex]
\quad formation && magnetic arcades && flux tubes \\
\hline
\end{tabular}
\end{table}
%%%%%%%%%%%%%%%%%%%%%%%%%%%%%%%%%%%%%%%%%%%%%%%%%%

In fact, the idea of splitting was outlined already in the first work on
the anemone microflares \citep[see Figures~3(D,E) in][]{Shibata_07}.
However, as far as we know, the first attempt of the quantitative description
of such a splitting was undertaken only in our recent paper \citep{Dumin_19}.
As a working tool, we employed the so-called GKSS model of the magnetic field
(\opencite{Gorbachev_88a}; for mathematical details, see also
\opencite{Brown_99}; \opencite{Somov_13}).
Its most important feature is a kind of the ``topological instability'',
namely, a sudden emergence (due to the bifurcation) of the additional
null point high above the plane of the sources; and the position of this point
(and, therefore, the pattern of splitting of the magnetic fluxes) drastically
depends on tiny displacement of the sources.%
\footnote{%
We called this instability topological because, from the mathematical
point of view, it is associated with change in the topological indices of the
vector field \citep{Gorbachev_88a}.
In other words, this is the instability of topology of the magnetic field.
This issue will be discussed in more detail in a separate paper.}

In the simplest version, the above-mentioned model involves four magnetic
sources (sunspots)---two positive and two negative---located in the same plane.
The global 3D pattern of the magnetic-field lines is given by the so-called
two-dome structure, where four domains of different topological connectivity
are shaped by the two superimposed domes (separatrices), which intersect each
other along the separator.
The primary energy-release region is assumed to be somewhere near the top
of the separator (it is presumably caused by the magnetic reconnection due to
the local current systems, but we do not discuss here the corresponding
processes in more detail.)
Next, we can consider a set of the magnetic field lines emanating from
various points of the energy-release region and propagating in space up to
their intersection with the plane of the chromosphere, located somewhat above
the plane of the sources
\citep[for more details, see Figure~3 in][]{Dumin_19}.%
\footnote{%
In fact, this idea was put forward long time ago for the interpretation of
ordinary two-ribbon flares by \citet{Gorbachev_88b}.}

\begin{figure}
\centerline{\includegraphics[width=0.62\textwidth]{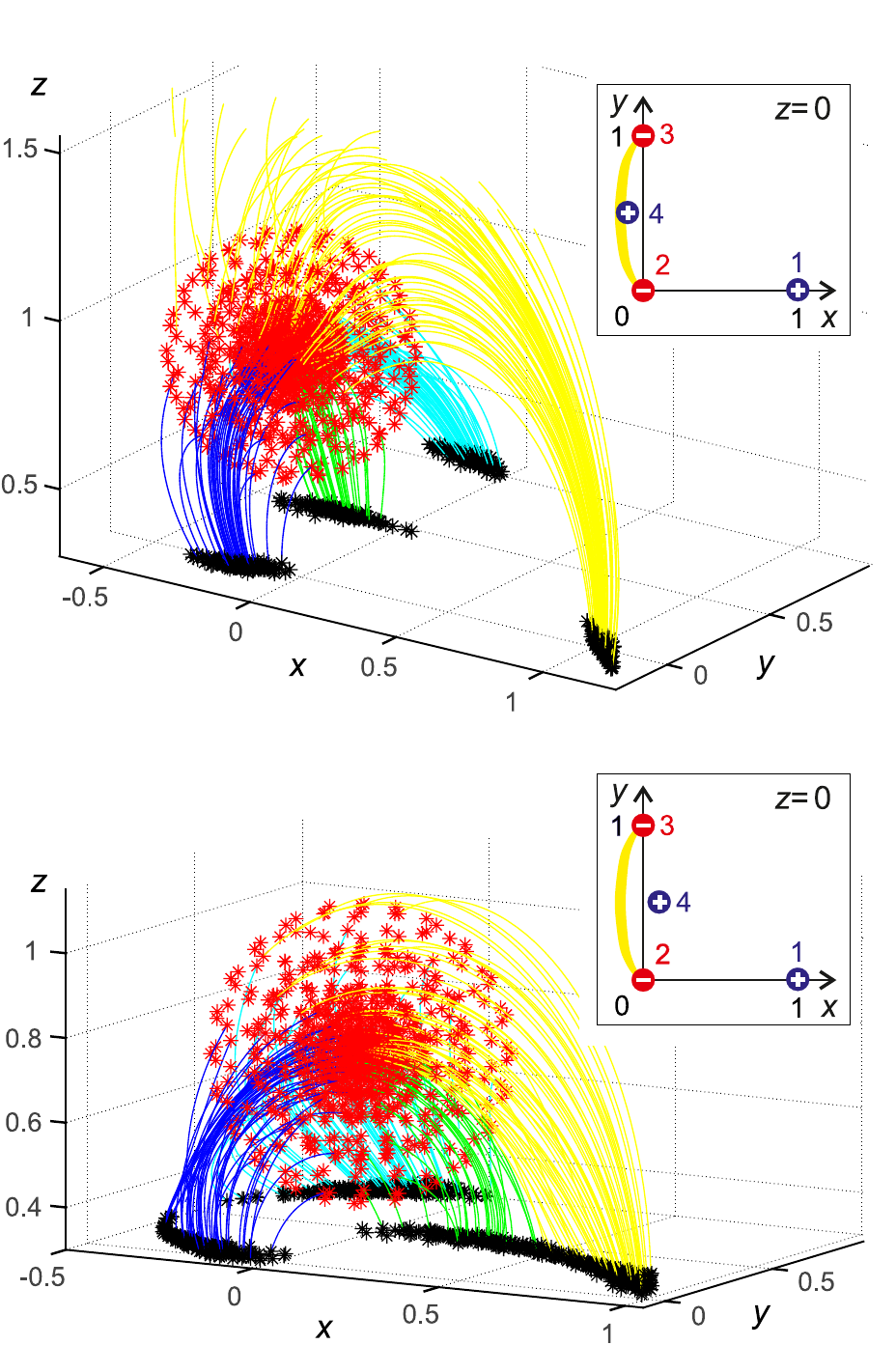}}
\caption{
Splitting of the magnetic-flux tubes originating from the energy-release
region for the magnetic-source arrangement near the zone of topological
instability (top panel) and away from this zone (bottom panel).
Each red asterisk in the energy-release region is an initial point of
the magnetic field line, which crosses the plane of emission $ z\,{=}\,0.3 $
(in the dimensionless units) at the respective black asterisk.
The magnetic-field fluxes with different topological connectivity
(\textit{i.e.}, terminating at different magnetic sources/sinks) are shown
by different colors (yellow, green, blue, and cyan).
The insets to the both panels show arrangement of the point-like magnetic
sources with equal magnitudes---two positive (1 and 4) and two negative
(2 and 3)---in the plane $ z\,{=}\,0 $.
The yellow semilunar region in these insets is just the zone of topological
instability.}
\label{Fig1}
\end{figure}

Strictly speaking, four sources should always produce four topologically
different magnetic fluxes.
However, some of these fluxes can merge to each other and, as a result,
the number of the visible footprints will be less.
As follows from the detailed calculations, the fluxes form four distinct
emission spots only when the magnetic sources are located near the
above-mentioned narrow zone of topological instability (\textit{i.e.}, when
the fluxes can interconnect with each other in the bifurcated null point);
see top panel in Figure~\ref{Fig1}.%
\footnote{%
To avoid misunderstanding, let us emphasize that this figure is just
a sketch of splitting the energy fluxes, which is not immediately related
to the subsequent observational data.}
Otherwise, only three (or even two) emissive ribbons are formed;
\textit{e.g.}, bottom panel in Figure~\ref{Fig1}.
A representative set of various geometrical configurations of the emission
spots can be found in Figure~5 of our previous paper \citep{Dumin_19};
and they closely resemble the variety of AMFs reported previously in
the literature \citep{Shibata_07,Nishizuka_11,Singh_11,Singh_12}:
Namely, the most part of the flares have three ribbons, but there is also
a small fraction of AMFs (presumably, corresponding to the zone of topological
instability) possessing four ribbons.

%%%%%%%%%%%%%%%%%%%%%%%%%%%%%%%%%%%%%%%%
%%%%%%%%%%%%%%%%%%%%%%%%%%%%%%%%%%%%%%%%

The topological instability is realized, roughly speaking, when three
magnetic sources with intermittent polarity are situated approximately
along the same line, while the fourth source is located aside from them.
In other words, they are arranged in the T-like configuration; for more
details, see Figure~1 in our paper \citet{Dumin_19} and the corresponding
discussion there.
In general, it is a very non-trivial fact that the number of the emissive spots
is determined just by the zone of topological instability, because the concept
of this instability was originally introduced by \citet{Gorbachev_88a} in
absolutely different context: it was associated with bifurcation and subsequent
fast motion of a null point of the magnetic field, which might be responsible
for the fast magnetic reconnection \citep[\textit{e.g.},][]{Dumin_17}.
However, quite unexpectedly, the same bifurcation turned out to be also of
crucial importance for splitting/merging of the projections of the magnetic
fluxes.
Of course, GKSS model is only one of possible options for the configuration
of magnetic sources.
However, by varying a single free parameter, it reproduces surprisingly well
almost all typical structures of the AMF emissive ribbons.

A few natural questions arise now:
\begin{enumerate}
\item
Can the same kinds of three- and four-ribbon AMFs be formed in the
substantially different magnetic-source configurations (for example, with
an unbalanced magnetic polarity)?
\item
Is it possible to get more complex AMFs (namely, involving more than four
ribbons) when a more sophisticated combination of the magnetic sources
is realized?
\end{enumerate}
As far as we know, such cases were not reported in the previous literature
(apart from the ``atypical'' flares of much greater size); and it will be the
aim of the subsequent section to represent the corresponding observations.

\section{Observational Results}
\label{sec:Observ}

To answer the above-posed questions, we performed a visual inspection of
a large series of pictures of the emission in {Ca}\,{\small II}\,H line taken
by the Solar Optical Telescope (SOT) onboard \textit{Hinode} spacecraft
\citep{Kosugi_07,Tsuneta_08}, which are stored in the Hinode archive,%
\footnote{%
See \url{https://hinode.nao.ac.jp/en/for-researchers/qlmovies/top.html} .}
and confronted them with the corresponding magnetograms by the Helioseismic
and Magnetic Imager (HMI) onboard \textit{SDO} satellite \citep{Pesnell_12},
which are available at the Joint Science Operations Center (JSOC) web-site.%
\footnote{%
See \url{http://jsoc.stanford.edu/HMI/hmiimage.html} .}
We sought for the AMFs, first of all, in the regions of the solar surface
involving the complex configurations of the moderately-sized sunspots with
intermittent polarity.%
\footnote{%
It should be mentioned that authors of the early studies of AMFs
\citep{Shibata_07,Nishizuka_11,Singh_11,Singh_12} usually did not try
to confront the pictures of emission with the respective magnetograms.
Most probably, this was because the high-resolution images by \textit{SDO}/HMI
were unavailable at that time, while resolution of other magnetograms was
insufficient to judge about the spatial scales typical for AMFs.}
Since they develop predominantly in the periods of high solar activity, we have
analyzed in much detail, for example, the \textit{Hinode}'s data for 2014.
On the other hand, it was almost impossible to identify any AMFs in the years
of low activity.
There was no clear association of AMFs with the large-scale solar flares:
sometimes they occurred in the vicinity of the powerful flares, while often
beyond them.
At last, since we analyzed only on-disk observations, attention was focused
at the ribbons in the base of AMFs, while the corresponding chromospheric jets
were not taken into consideration.

%%%%%%%%%%%%%%%%%%%%%%%%%%%%%%%%%%%%%%%%
\begin{figure}[t]
\centerline{\includegraphics[width=0.7\textwidth]{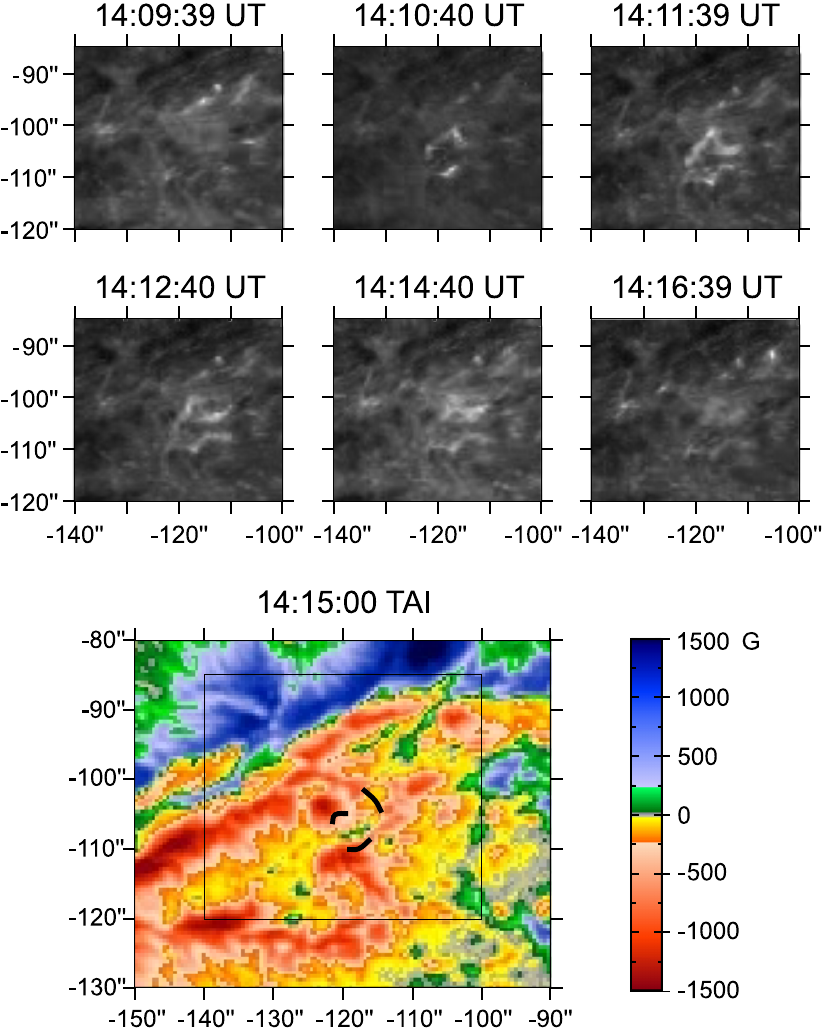}}
\caption{Example of the three-ribbon AMF on 2 February 2014 that
was formed by the unbalanced magnetic sources:
a temporal sequence of the emission patterns in
{Ca}\,{\small II}\,H line recorded by \textit{Hinode}/SOT (top panels)
\textit{vs.}\ the pattern of magnetic fields obtained by \textit{SDO}/HMI
(bottom panel, courtesy of NASA/SDO and the HMI science team).
A sketch of the ribbon configuration at 14:10:40~UT is plotted in the
magnetogram.}
\label{Fig2}
\end{figure}
%%%%%%%%%%%%%%%%%%%%%%%%%%%%%%%%%%%%%%%%

As in the previous study, we preferred to use {Ca}\,{\small II}\,H line
because it is emitted by a moderately heated plasma, $ {\approx}10^4 $~K.
So, one can expect that the fine spatial features in the emission patterns are
caused just by the geometric properties of the magnetic field lines rather
than by the ``physical'' filamentation (such as development of the plasma
instabilities, MHD waves, \textit{etc.}).

Let us mention that the coordinate frame of \textit{Hinode}/SOT experiences
a substantial instability in the east--west direction.
We did not try to compensate this instability but instead adjusted
the magnetogram's coordinates to the ones by \textit{Hinode}/SOT.
The color map in the magnetograms presented below looks rather unusual,
it conforms with the recommendations of HMI team at the above-mentioned JSOC
web-site (file \texttt{HMI{\_}M.ColorTable.pdf}): the weak fields
(within~$ \pm 24 $~G) are shaded in gray, and the colors change sharply
at~$ \pm 236 $~G, when the magnetic field begin to affect the photospheric
brightness (\textit{i.e.}, the sunspots are formed).
Such color coding was designed to visually show the structure at both high
and low field values.
As follows from our experience, this color coding is really convenient.

\subsection{AMF with Strongly Unbalanced Magnetic Fluxes}
\label{sec:AMF_unbal}

The above-mentioned analysis enabled us to identify some interesting cases of
the AMFs, unknown before.
The first non-trivial situation, observed on 2~February 2014, is shown in
Figure~\ref{Fig2}.
Its emission patterns look like in the ordinary three-ribbon AMF:
the ribbons quickly develop at the time scale about 1~min and then become
diffuse and gradually decay in the subsequent few minutes.
However, the corresponding magnetic sources are evidently unbalanced:
The magnetogram demonstrates a complex cross-like arrangement of a few
negative sources; but the positive sources exist only at very large distances,
mostly in the northern part of the picture (while, according to the standard
GKSS model, there must be at least one positive source in the vicinity of
the ribbons).
So, to explain such flares, GKSS model should be generalized to the case of
unbalanced magnetic fluxes.

%%%%%%%%%%%%%%%%%%%%%%%%%%%%%%%%%%%%%%%%
\begin{figure}[t]
\centerline{\includegraphics[width=0.67\textwidth]{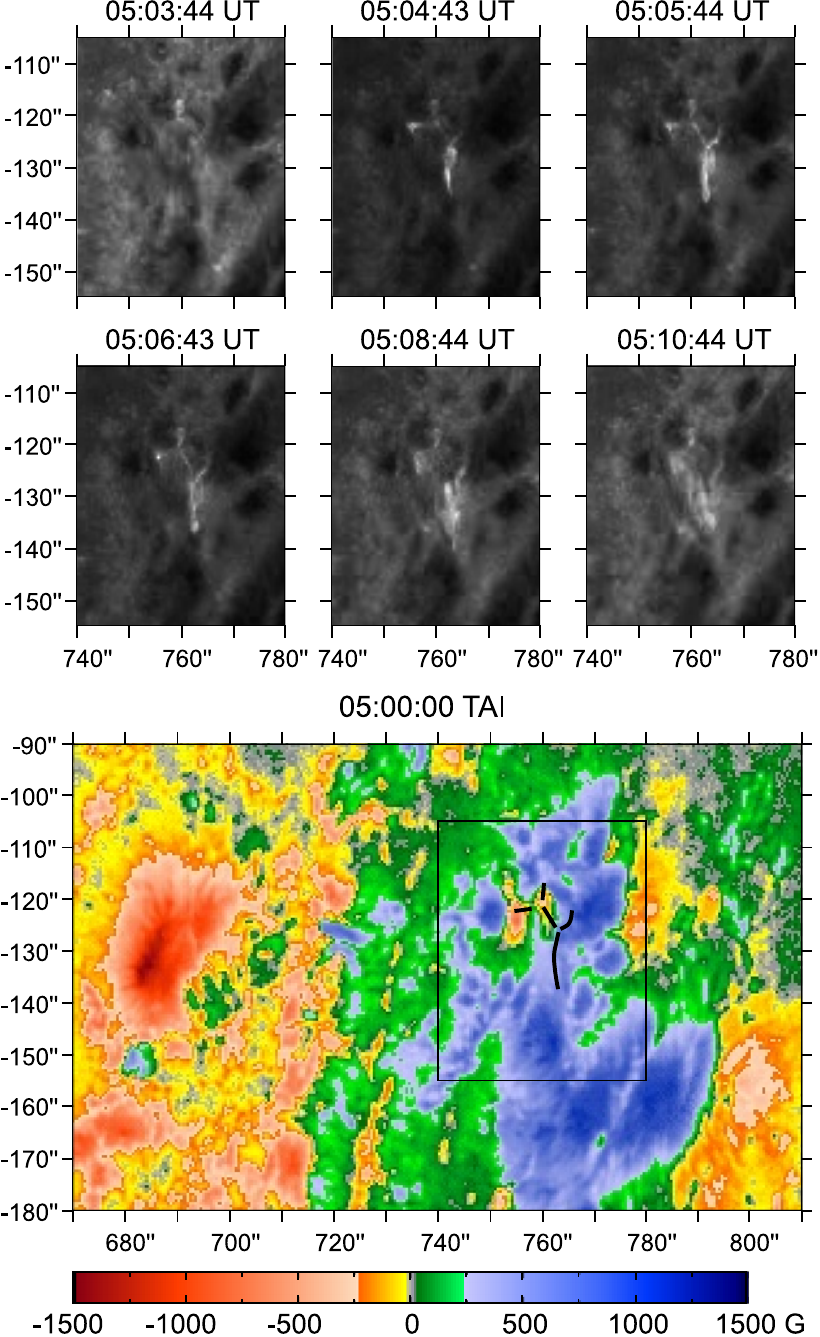}}
\caption{Example of the five-ribbon flare in the solar chromosphere on
16~February 2014: a temporal sequence of the emission patterns in
{Ca}\,{\small II}\,H line recorded by \textit{Hinode}/SOT (top panels)
\textit{vs.}\ the map of magnetic fields obtained by \textit{SDO}/HMI (bottom
panel, courtesy of NASA/SDO and the HMI science team).
A sketch of the ribbon configuration at 05:05:44~UT is plotted in
the magnetogram.}
\label{Fig3}
\end{figure}
%%%%%%%%%%%%%%%%%%%%%%%%%%%%%%%%%%%%%%%%

\subsection{AMF with Numerous Ribbons}
\label{sec:AMF_ribbons}

The second---and most interesting---example of the unusual AMF is shown in
Figure~\ref{Fig3}: this is a flare with a larger number of the emissive ribbons
than in the previously-known three- and four-ribbon cases.
It was recorded on 16~February 2014 approximately at 05~UT.
As is seen, the flare begins to develop at 05:04:43~UT with three or four
ribbons (depending on their definition).
Then, at 05:05:44 and 05:06:43~UT one can see a complex structure composed
of five (or even six) ribbons.
At last, starting from 05:08:44~UT, the ribbons acquire a diffuse character
and gradually disappear.

At the first glance, a semi-circular segment in the bottom right side of this
structure looks like an arc extending into the upper layers of the atmosphere.
However, one should keep in mind that this AMF occurred quite close to
the west limb.
Then, a projection of such an arc onto the picture plane would be bent to
the limb, while just the opposite behavior is seen in the figure.
So, we should conclude that the above-mentioned segment is located most
probably on the surface.

A lifetime of the entire structure was 2--3~minutes, which is comparable to
(or even somewhat less than) the lifetimes of ordinary AMFs reported before
\citep[\textit{e.g.}, Figure~2 in][]{Shibata_07}.
On the other hand, a characteristic length of the entire structure in this case
was about $ 20'' $ (15\,000~km), which is substantially greater than in
the ordinary three- and four-ribbon AMFs ($ 3''{-}7'' $, or 2000--5000~km).
Besides, while thickness of the ribbons in the ordinary AMFs was
$ 0.2''{-}0.4'' $ (150--300~km), in our case they can be much thicker
(up to $ 1''{-}2'' $, or 700--1500~km); but this depends on the particular
ribbon, and some of the ribbons remain thin.

To understand the structure of magnetic field responsible for this AMF,
we superimposed a sketch of the emissive ribbons onto the magnetogram of
the corresponding region (bottom panel in Figure~\ref{Fig3}).
As is seen, the flare is produced by four magnetic sources (positive, negative,
negative, and again positive) along the parallel and a few sources (mostly,
positive) away from them, along the meridian.
Such a configuration qualitatively reminds GKSS model of the magnetic field,
where the topological instability with a complex flux splitting takes place,
roughly speaking, at the T-like arrangement of the magnetic sources
(see inserts in Figure~\ref{Fig1});
but this case evidently involves a larger number of the sources.%
\footnote{%
One of the theoretical models with a large number of magnetic sources was
developed long time ago by \citet{Inverarity_99}; but these sources had a very
symmetric arrangement, which can be hardly relevant to the realistic AMFs.}
A detailed topological analysis of such magnetic configurations is still
to be done \citep[for the recent reviews of topological models of the solar
magnetic fields, see][]{Longcope_05,Janvier_17}.
Besides, the total magnetic flux in the vicinity of AMF again seems to be
substantially unbalanced, as in Figure~\ref{Fig2}.

In fact, the AMF presented in Figure~\ref{Fig3} looks like a combination of two
three-ribbon flares possessing one common ribbon.
We observed also a few other cases when two three-ribbon structures were
separated by a larger distance and did not have a common ribbon, but they
appeared almost simultaneously (within 1--2~minutes).
So, it is difficult to say if that was a single AMF or two different flares
developing at the same time (in other words, if they were associated with one
or two independent systems of the split magnetic fluxes and the corresponding
energy-release regions)?
It is interesting to mention that a ``recurrent'' (two-step) chromospheric
microflare was observed also by \citet{Zeng_16}, and the corresponding
``steps'' were separated by the time interval as long as 20~minutes.

In general, the multi-ribbon structures seen in Figures~5(a,b) from the
above-cited work are very similar to AMFs.
However, they were observed in the line He\,{\small I}\,10\,830\,{\AA}, which
is sensitive to the higher temperatures than {Ca}\,{\small II}\,H.
So, these images trace mostly the magnetic field lines themselves rather than
their projections (ribbons) onto the denser layers.

\section{Theoretical Interpretation}
\label{sec:Theory}

It is a very nontrivial task to understand the geometric configuration of
the reconnecting field lines in the unusual AMFs reported above.
Really, a generic structure of the 3D null point, responsible for the
magnetic reconnection, involves six asymptotic directions at $ {\bf r} \to 0 $
(where $ {\bf r} = 0 $ is the null point), along which the magnetic field
lines originate or disappear \citep{Parnell_96,Dumin_16}.
Four of these asymptotes are ``dominant'', so that the corresponding fluxes
are concentrated and terminate at the magnetic sources of different polarity
(two positive and two negative).
Two other asymptotes are ``recessive'', so that the respective field lines
quickly deviate from them with increasing distance and scatter in space.
Therefore, a standard pattern of the magnetic reconnection involves four
fluxes; two of them being rooted in the positive sources, and two in
negative.

So, if there are more than four fluxes in total or more than two fluxes of
the same type, the standard picture is not applicable.
Then, we should assume that (i) either there are more than one null point
in the region of reconnection or (ii) the reconnection occurs in the
high-order null point, where the number of incoming and outgoing fluxes
can be greater \citep{Zhugzhda_66,Lukashenko_15,Lukashenko_16}.

However, from the mathematical point of view, the case of the high-order
null point is degenerate, and the probability of its realization tends to
zero.
So, the assumption of a few closely-located null points in the region of
reconnection looks preferable.
This point of view is especially supported by the recent data by
\citet{Zeng_16}, who clearly observed a ``sequential branching'' of the
magnetic-flux tubes; see Figures~3(a) and 5(a,b) in the last-cited paper.

\section{Conclusions}
\label{sec:Concl}

\begin{enumerate}
\item
The anemone microflares in the solar chromosphere represent a specific type of
the transient phenomena, which are substantially different from the ``atypical''
multi-ribbon flares both from the viewpoint of their lifetime and the spatial
scales.
The AMFs, observed in the chromospheric lines, are not the ``isolated''
phenomena but manifest themselves also in EUV an X-rays, as was found recently
by \citet{Zeng_16}.
Most probably, the complex multi-ribbon structure of AMFs is formed due to
magnetic flux splitting at some height above the ribbons.
\item
The GKSS model of magnetic field, utilized in our previous paper
\citep{Dumin_19}, can give a rather good statistical description for the
majority of three- and four-ribbon AMFs.
However, as follows from the present work, it is possible to observe sometimes
the more exotic flares, which are evidently beyond this model.
These are, for example, the AMFs with more than four emissive ribbons as well
as the flares occurring in the regions with strongly unbalanced magnetic
polarity.
Most probably, such flares involve a few null points in the region of
reconnection, resulting in the sequential branching of the magnetic fluxes.
However, the detailed topological models of these events still need to be
developed.
\end{enumerate}

\begin{acks}

YVD is grateful to A.V.~Getling, A.V.~Oreshina, and I.V.~Oreshina for
consultations on the processing of magnetic fields,
A.T.~Lukashenko for the discussion of the high-order null points,
as well as to G.G.~Motorina and A.V.~Stepanov for valuable comments and
suggestions.

\textit{Hinode} is a Japanese mission developed and launched by ISAS/JAXA,
with NAOJ as domestic partner and NASA and STFC (UK) as international partners.
It is operated by these agencies in co-operation with ESA and NSC (Norway).

\smallskip
\noindent {\bf Declaration of Potential Conflicts of Interest} \quad
The authors declare that they have no conflict of interest.

\end{acks}

\bibliographystyle{spr-mp-sola}
% \bibliography{Dumin_Somov}

\end{article} 

\end{document}